\documentclass[a4paper,11pt]{article}
\usepackage{pos}

\title{
  \vspace{-1.2cm}
Tests and characterisation of the KI trigger for fast events on the EUSO-SPB2 Fluorescence Telescope
  \vspace{-0.3cm}
}
 \ShortTitle{EUSOSPB2 KI TRIGGER}

\author*[a,b]{Hiroko Miyamoto}
\author[b,f]{Matteo Battisti}
\author[c,d]{Alexander Belov}
\author[b,e]{Mario Bertaina}
\author[f]{Sylvie Blin}
\author[f]{Alxandre Creusot}
\author[g]{Johannes Eser}
\author[h]{George Filippatos}
\author[d]{Pavel Klimov}
\author[b,e]{Massimiliano Manfrin}
\author[b]{Marco Mignone}
\author[f]{Etienne Parizot}
\author[i]{Lech Wiktor Piotrowski}
\author[f]{Guillaume Pr\'{e}v\^{o}t}
\author[c,d,f]{Daniil Trofimov}

\affiliation[a]{Gran Sasso Science Institute, 
  Viale Francesco Crispi 7, 67100 l'Aquila, Italy
}

\affiliation[b]{INFN Section of Turin, 
  Via P. Giuria 1, 10125 Turin, Italy
}

\affiliation[c]{M.V.Lomonosov Moscow State Univ., Faculty of Physics, 
  Leninskie gory 1(2), 119234 Moscow, Russia
}

\affiliation[d]{M.V.Lomonosov Moscow State Univ., Skobeltsyn Institute of Nuclear Physics, 
  Leninskie gory 1(2), 119234 Moscow, Russia
}

\affiliation[e]{Univ. of Turin, Dept. of Physics, 
  Via P. Giuria 1, 10125 Turin, Italy
}

\affiliation[f]{Univ. Paris Cit\'e, CNRS, Astroparticule et Cosmologie, 
  10 Rue Alice Domon et \`{e}onie Duquet, 75013 Paris, France
}

\affiliation[g]{The Univ. of Chicago, Dept. of Astronomy and Astrophysics, 
  5640 S. Ellis Avenue, 60637 Chicago IL, US
}

\affiliation[h]{Colorado School of Mines, Dept of Physics, 
  1523 Illinois St., 80401 Golden CO, US
}

\affiliation[i]{Univ. of Warsaw, Faculty of Physics, 
  Ludwika Pasteura 5, 02-093 Warsaw, Poland
}

\onbehalf{\\for the JEM-EUSO collaboration}

\emailAdd{miyamoto@to.infn.it}
\emailAdd{matteo.battisti@to.infn.it}
\emailAdd{parizot@apc.in2p3.fr}
\emailAdd{trofimov@apc.in2p3.fr}

\abstract{

  The second generation Extreme Universe Space Observatory on a Super-Pressure Balloon (EUSO-SPB2) mission is a stratospheric balloon mission developed within the Joint Exploratory Missions for
  Extreme Universe Space Observatory (JEM-EUSO) program. The Fluorescence Telescope (FT) is one of the two separate Schmidt telescopes of EUSO-SPB2, which aims at measuring the fluorescence emission of extensive air showers from cosmic rays above the energy of 1 $EeV$, looking downwards onto the atmosphere from the float altitude of 33 km. The FT measures photons with a time resolution of 1.05 ${\mu}$s in two different modes: single photon counting (PC) and charge integration (KI).  In this paper, we describe the latter and report on the measurements of its characteristics. We also present a new trigger based on this channel, the so-called KI trigger, which allows to measure additional types of events, namely very short and intense light pulses. We report on the tests of this trigger mode in the laboratory and at the TurLab facility, and its implementation in the EUSO-SPB2 mission.
  }

\ConferenceLogo{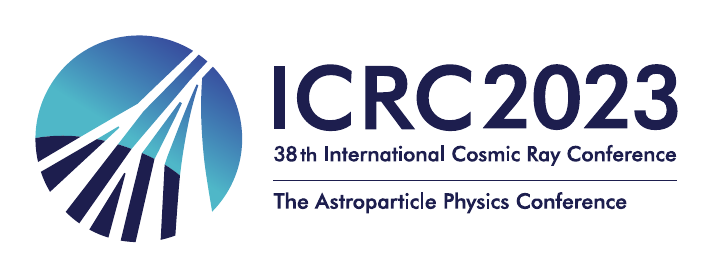}

\FullConference{%
38th International Cosmic Ray Conference (ICRC2023)\\
  26 July - 3 August, 2023\\
  Nagoya, Japan}

\begin{document}
\maketitle
\vspace{-0.2cm}
\section{Introduction} 
\vspace{-0.2cm}
The second generation Extreme Universe Space Observatory on a Super-Pressure Balloon (EUSO-SPB2)
mission~\cite{ref:spb2} is a stratospheric balloon mission developed within the Joint Exploratory Missions for Extreme Universe Space Observatory (JEM-EUSO) program~\cite{ref:jemeuso}. It hosts two main telescopes: a fluorescence telescope (FT), looking towards the nadir and aiming primarily at the detection of extensive air showers (EAS) from high-energy cosmic rays (above $\sim 10^{18}$~eV) \cite{ref:spb2}, and a Cherenkov telescope (CT), looking almost horizontally within a few degrees above and below the limb of the Earth, aiming at the detection of Cherenkov light from air showers induced by cosmic rays (above the limb) or upward going neutrinos (below the limb) ~\cite{ref:ANITA}.

The FT is composed of 3 photo-detection modules (PDMs), each of which features 9 elementary cells (ECs) composed of 4 Hamamatsu 64-ch Multi-Anode-Photomultiplier Tubes (MAPMTs), for a total of 2304 pixels. An EC is a compact assembly of HVPS generator board and front-end electronics based on a SPACIROC-3 ASIC~\cite{ref:SPACIROC3} for each PMTs, potted in a gelatinous compound to prevent discharge between the various components.

\vspace{-0.2cm}
\section{Photon counting and charge integration}
\vspace{-0.2cm}

\begin{figure}[t]
  \centering
  \includegraphics[width=15cm,clip]{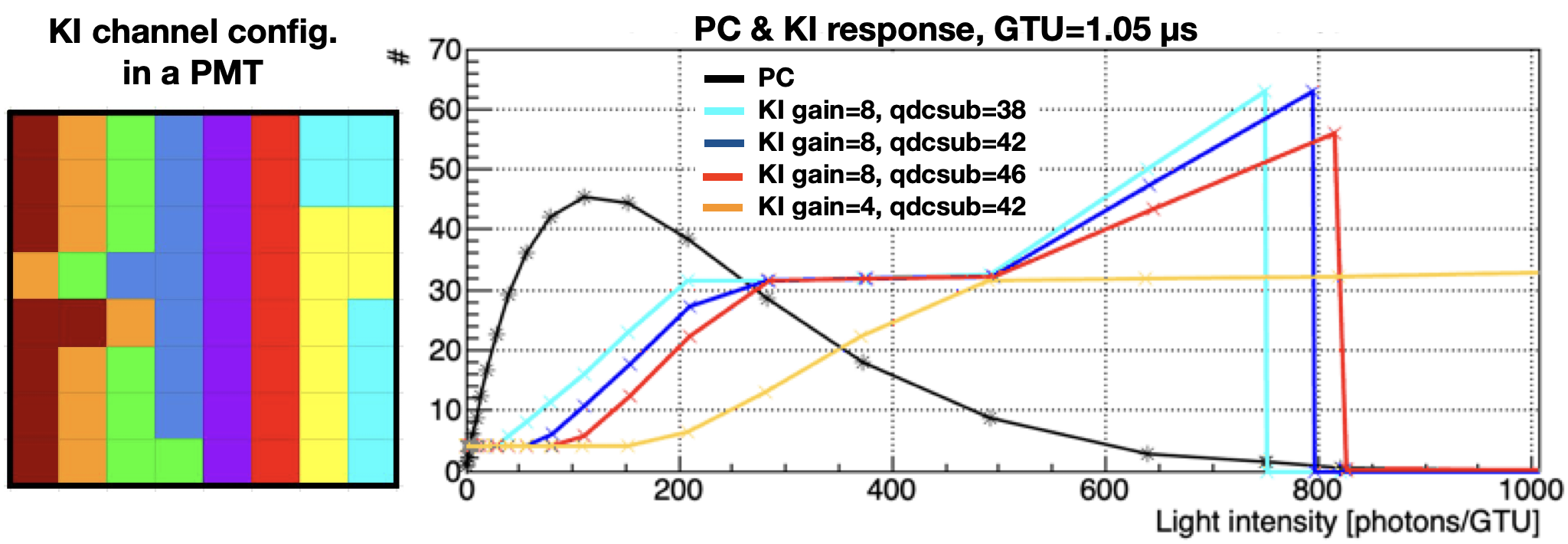}
  \caption{
    Left: distribution of the KI channels on a PMT. Pixels with the same colour belong to the same KI channel.
    Right: Photon counts per GTU (in black) and KI values as a function of light intensity, in number of photons per GTU. The curves correspond to a KI gain of 8 and values of qdcsub of 38 (light blue), 42 (blue), 46 (magenta) and KI gain of 4 and qdcsub of 42 (yellow).
    \vspace{-0.3cm}
  }
  \label{fig:KImap}
  \end{figure}

The so-called Gate Time Unit (GTU) is the basic time window of the FT, which is set to 1.05$\mu$s in EUSO-SPB2. Over each GTU, the ASIC performs two types of measurements: photon counting (PC) and charge integration (referred to as ``KI''). The PC functionality allows the independent counting of charge signals on each of the 64 pixels of the MAPMT, with a time resolution of $\delta t \simeq 6$~ns. The anode signals go through a low input impedance preamplifier followed by a fast discriminator, optimized for single photon detection. The threshold is set by a common 10-bit DAC register for each PMT, as well as a 7-bit DAC for each individual channel, used to compensate the MAPMT pixel gain dispersion. When two photoelectrons are generated at the photocathode within an interval of time lower than the resolution $\delta t$, only one photon is counted. Therefore, for light intensities corresponding to more than $\sim 20$--25 photoelectrons, pile-up~\cite{ref:pileup} becomes important and the photon counting is not anymore proportional to the actual photon flux. An example of the resulting saturation curve is shown in the right panel of Fig.~\ref{fig:KImap}, where the photon counts are seen to first increase linearly with the photon intensity, then reach a maximum around 45 per GTU, and finally decrease to zero as more and more photons are ``piled-up'' together.

To 
help solving the ambiguity on the true number of photon counts in case of large photon counts, 
the KI channels of the SPACIROC-3 ASIC can be used. There are 8 KI channels per ASIC (thus, per MAPMT), 
each of which gathers 8 pixels of an MAPMT, 
according to the patterns as shown in the left panel of Fig.~\ref{fig:KImap}
(whose shapes are dictated by tight routing constraints on the ASIC board inserting inside the EC). These channels perform an integration of the charge collected at the 8 anodes of the corresponding pixels, which is then read through a time of discharge and coded on 6 bits (i.e. with values from 0 to 63). Although the response of the KI channel is not linear with the photon flux (due to internal ASIC characteristics as well as limitations of the power supply inherent to the very low consumption system), it allows a clear discrimination between low values of the integrated charge (for which the KI returns its default value, namely 4), intermediate signals for which the KI value increases from 4 to 63 (with a plateau spreading over a large part of this range), and very large signals, namely too large to be discharged with one GTU, for which the KI then returns the value 0. An example of the KI response as a function of light intensity can be seen in the right panel of Fig.~\ref{fig:KImap}, in the case of a constant intensity.
Each individual channel can be configured with 2 parameters: a gain (amplifying or reducing the input charge signals) and a subtraction current (to compensate the internal DC current associated with the channel). Increasing the gain results in a shift of the KI response curve to the left, while increasing the subtraction current (adjusted through the so-called ``qdcsub'' parameter) results in a shift to the right.

An important application of the KI measurements in the framework of EUSO-SPB2 is to allow the detection of very short light pulses, of the order of or lower than the time resolution of the PC channel. 
Such a signal would be essentially counted just 1 in PC mode independent from the charge, while it could be measured by its intensity through their corresponding KI signal.
Such short, intense signals are for instance expected from the Cherenkov emission of EAS induced by cosmic rays when the FT camera is pointed towards the sky from the ground (as during the field tests of EUSO-SPB2). In flight, with nadir pointing, speculative upward-going events on nanosecond scale, similar to those reported by ANITA (with steeper emergence angles), could also give rise to clear KI signals~\cite{ref:ANITA}. Therefore, it is interesting to extend the triggering scheme originally developed in the framework of the previous balloon flight EUSO-SPB1~\cite{ref:spb1}, to use the KI signals in addition to the photon counting.

\vspace{-0.1cm}
\section{KI trigger}

\vspace{-0.1cm}

\begin{figure}[t]
  \centering
  \vspace{-0.4cm}
  \includegraphics[width=11cm,height=6.4cm]{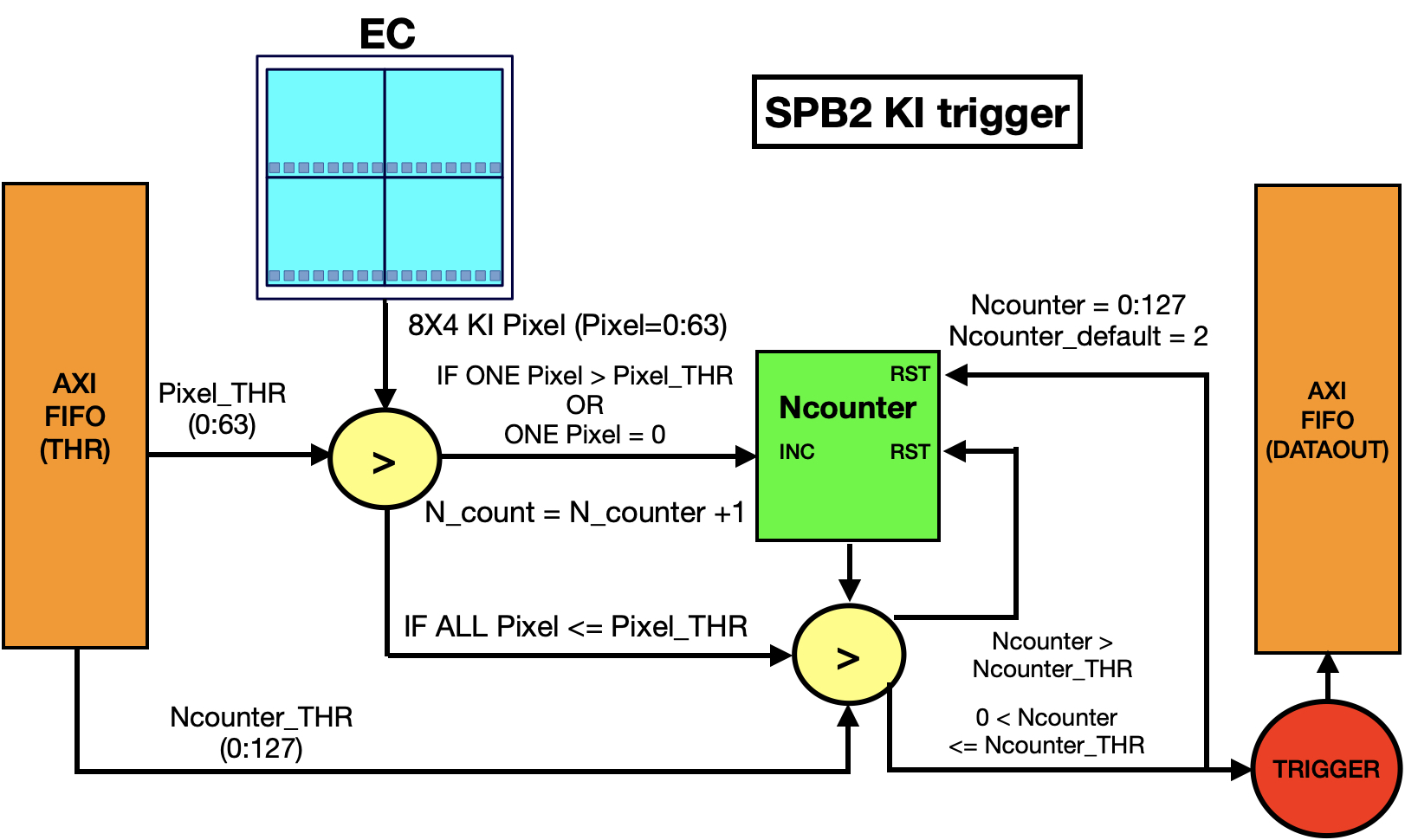}
  \caption{
    \vspace{-0.2cm}
    SPB2-KI trigger logic
    \vspace{-0.4cm}
  }
  \label{fig:KItrigger}
  \end{figure}

For EUSO-SPB2, two
types of triggers have been developed: a Level 1 (L1) trigger and a KI trigger. The FT operation system allows to work either with only any one of them, or with both in parallel. The logics of the L1 trigger has been described elsewhere~\cite{ref:SPB2L1}. Here, we concentrate on the KI trigger logic, of which a schematic view is shown in Fig.~\ref{fig:KItrigger}.

The KI trigger has two key parameters: a threshold, Pixel\_THR, and a time duration, in numbers of GTUs, Ncounter. A trigger is issued when a KI channel records a value larger than the threshold, for a number of GTUs \emph{lower} (not larger!) than the predefined duration. The idea behind it is to concentrate on fast signals (pulses or events lasting less than Ncounter GTUs), and not allow mere bright areas (or occasionally defective pixels) to trigger the system.

More precisely, when 
one KI channel value exceeds the threshold value Pixel\_THR, or is equal to 0 
(even larger signal), 
the counter [Ncounter] is incremented. If the signal does not last more than Ncounter GTUs (in the same EC), a trigger is issued. 
If the signal lasts more than Ncounter GTUs above the threshold (or with KI $= 0$), the counter is reset and does not trigger.

During the flight of EUSO-SPB2, the Ncounter parameter was set to 2, while Pixel\_THR was given different values at different periods, between 10 and 20.

\vspace{-0.1cm}
\subsection*{KI response to light pulses}
\vspace{-0.1cm}

The above-mentioned KI response
to 
light (see the right panel of Fig.~\ref{fig:KImap}) corresponds to the case of continuous light. In the case of a very short pulse, unfortunately, the situation is more complicated, for two independent reasons: one related mostly to the ASIC, the other to the HVPS. 

Regarding the ASIC, if the subtraction current parameter (``qdcsub'') is not perfectly adjusted to exactly compensate the DC current associated with a particular KI channel, then the actual charge measured is not exactly the charge that has been received at the 8 anodes of the MAPMT connected to that KI channel. If the subtraction is too low, then some charge is building up, even in the absence of actual light, and the KI values are overestimated. On the other hand, if the subtraction is too large, even actual signal may be subtracted away, and the KI values are underestimated, compared to what would be obtained with perfect compensation. Moreover, in the latter case, the KI value obtained from a given light pulse will be different, depending on when the pulse arrived within the GTU. It will be lower (larger) if the pulse arrive earlier (later), because the overcompensation of its signal will have started earlier (later). The other difficulty is related to the fact that a very bright pulse of light will result in the production of many photoelectrons at once, which will then multiply through the MAPMT up to a very large number of electrons in a very short time (a few ns). This will draw current at a level which cannot be instantly matched by the powering system. As a result, the gain will drop, which will lead to saturation. As a consequence, very bright and short pulses of different intensities (even by one order of magnitude) can lead to the same amount of charges entering the ASIC, and thus be integrated to produce the same value of KI. In addition, an oscillation of the electronics baseline results at the level the PMTs or even the EC, which can lead to fake, nearly uniform photon counts (typically between 1 and 4) if the PC threshold is set too close to the pedestal.

Because of the above features, with the current hardware the KI channels cannot be used to determine precisely the intensity of the incoming light. Nevertheless, they can indeed be used as clear indicators of intense light pulse, which would remain unnoticeable through the PC channel. This is the property that is exploited by the KI trigger. In addition, despite the above-mentioned saturation effect (which will be discussed and quantitatively studied in more detail elsewhere), the monotonous (though not proportional) relation between the light intensity and the KI values allows to derive comparative estimates of the detected signals.

\vspace{-0.2cm}
\subsection*{KI measurements inside blackboxes}
\vspace{-0.2cm}

KI measurements were made using pulsed LED inside blackboxes with two different setups, one at the Physics department of the University of Turin (Italy) and the other at APC, University of Paris (France). In both cases, the measurements were made using the same hardware as EUSO-SPB2, including an EC, the Zynq board and the HVPS. In Paris, we measured the KI response to an LED producing light pulses of 8~ns duration with an adjustable intensity, from a few hundreds to a few thousands of hundreds of photons, initially measured at high frequency with a calibrated NIST. The pulses were shot with a low frequency of 1~Hz and sent directly to a single pixel using an optical fibre. In Turin, the light from a sub-nanosecond LED pulse was focused on one pixel from a distance of $\sim 80$~cm, using an optical system with an Iris, a collimator and two plano-convex lenses. The same light was also measured by a SiPM whose output signal was recorded by a digital oscilloscope to estimate the approximate number of incident photons as a function of the intensity level of the sub-ns pulses. This setup allows additional light to illuminate the entire EC continuously, to simulate different levels of background light (similar to those encountered in flight) and verify their influence.

\begin{figure}[t]
  \centering
  \includegraphics[width=15cm,clip]{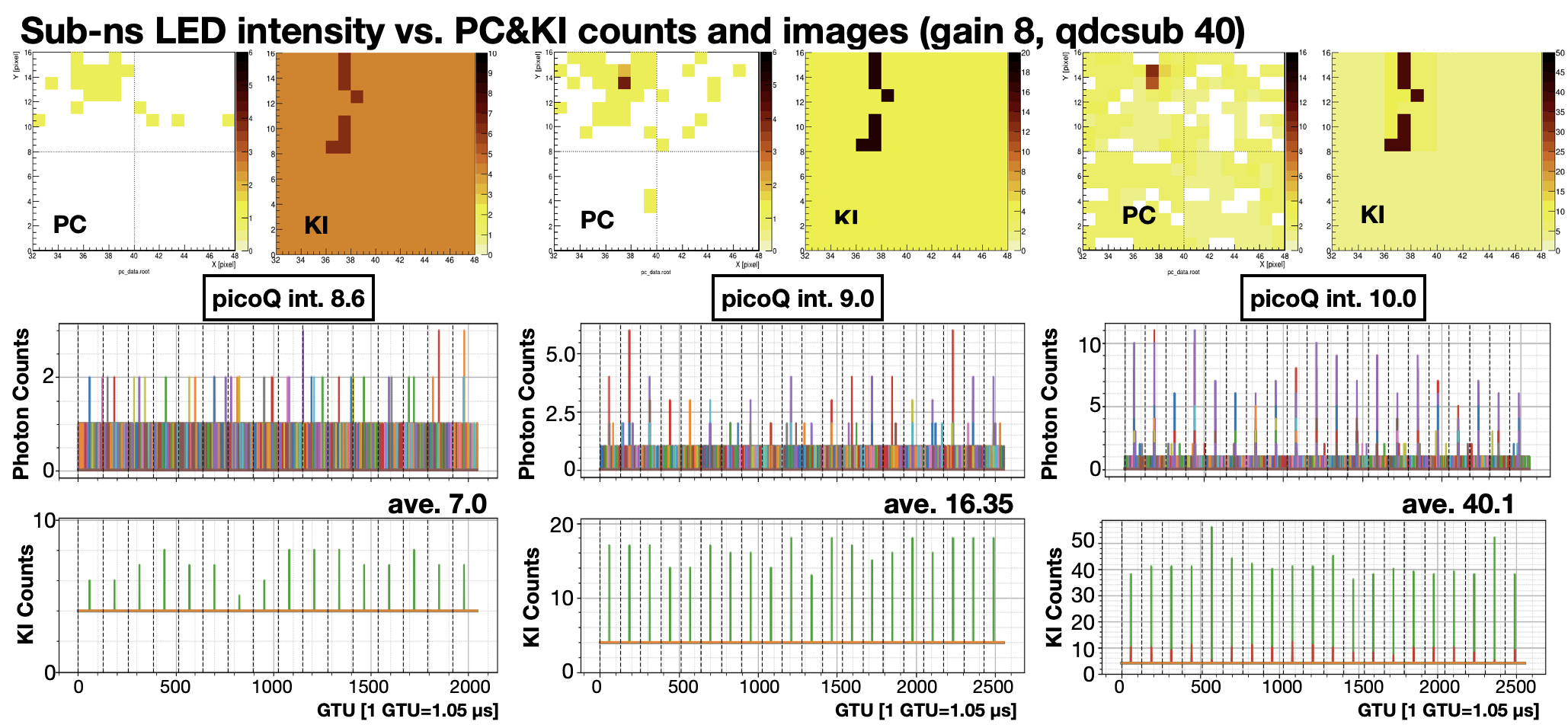}
  \caption{
    KI measurements in the blackbox for a different intensity of sub-ns pulsed LED light.
    Top: images of PC (left) and KI (right) of the LED focused light in a GTU.
    Bottom: PC (top) and KI (bottom) counts as a function of GTU in the LED nominal intensities of 8.6, 9.0 and 10.0. 
    \vspace{-0.3cm}
  }
  \label{fig:KIvsPC}
  \end{figure}

In Fig.~\ref{fig:KIvsPC}, we show the PC and KI response of the EC to a bright sub-ns pulse for different light intensities: 8.6 (left), 9.0 (center) and 10.0 (right), which correspond respectively to $\sim 4000$, $\sim 7000$ and $\sim 15000$ photons per pulse. The top line shows the PC and KI maps for each case, in a single GTU corresponding to when the pulse is produced. The middle line shows the photon counts as a function of time (in GTUs), for a periodic sequence of light pulses. The bottom line shows the corresponding KI response. 

As expected, the KI values are larger for larger light intensities, which validates the above-mentioned semi-quantitative use of the KI channels. For the 3 intensities, the typical KI values are round 7, 17 and 40, respectively. Individual values can however vary to some extent, which is reminiscent of the above remark regarding an imperfect DC current compensation and its effect on the KI values depending on the occurrence of the pulse within a GTU. It can also be seen that the photon counts remain low, around 2 or 3 at the lowest intensity, and up to between 5 and 10 at the largest one, as a result of the discussed baseline oscillation, which is indeed longer and thus allows for more (fake) photon counts in the case of a very strong current draw.

\subsection*{KI trigger tests at TurLab}

\begin{figure}[t!]
  \centering
  \includegraphics[width=14cm,clip]{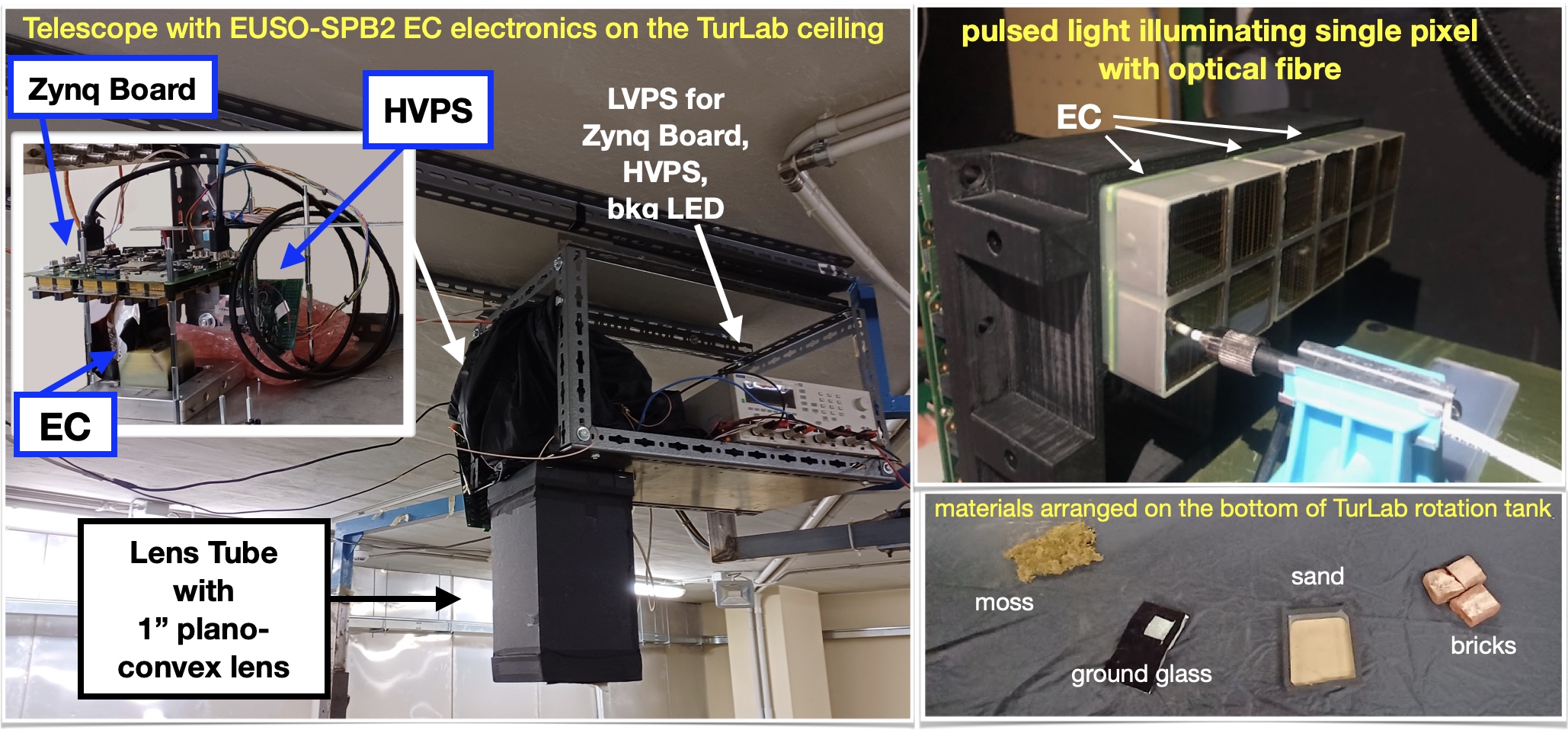}
  \caption{
    Left: a telescope with EUSO-SPB2 detector and electronics hanging on the ceiling above the TurLab rotation tank. 
    Bottom-right: examples of materials arranged 
    in the tank.
    Right (top): a photo of the 
    setup in the blackbox, where pulsed light is illuminating single pixel with optical fibre.
    \vspace{-0.5cm}
  }
  \label{fig:SPB2ECtel}
\end{figure}

The TurLab
is a laboratory for geo-fluid dynamics studies, where rotation is a key parameter such control the Coriolis force and Rossby Number.
The tank has a 5~m diameter and can rotate with periods from 1 to 20 min per rotation. It is located underground at the fourth basement of the Physics department building of the University of Turin, where the level of light can be controlled from almost complete darkness to any desired ambient brightness, to reproduce different conditions that can be encountered in flight. In practice, we place the detector under study at the focal surface of a small telescope attached to the ceiling above the tank, looking down to the the bottom of the tank where several materials and light sources are arranged. The rotation of the tank then allows to have these artefacts cross the field-of-view of the detector and reproduce variations of the background light and dynamical occurrences of various types of events, mimicking what the instrument can be expected to see during a flight. The dynamical response of several JEM-EUSO detectors have already been studied in this way in the framework of EUSO@TurLab project~\cite{ref:TurLab}, notably to test their performances and triggers (see e.g.~\cite{ref:SPB2L1}).

In the specific case of the KI trigger tests at the TurLab facility, we use an EC 
with the EUSO-SPB2 Zynq board and HVPS, assembled with a 1'' plano-convex lens tube in a telescope, as shown in Fig~\ref{fig:SPB2ECtel}.
At the bottom of the tank, crossing the field of view during tank rotation, we place various materials with different albedos (e.g. bricks, sand, ground glass and moss, mimicking rocks, desert, glacier, forest), as well as light sources such as the above-mentioned sub-ns pulsed LED (simulating very short and bright events), a strip of 10 white LEDs driven by an Arduino circuit (where each LED is pulsed sequentially for 5 ${\mu}s$, overlapping for 1${\mu}s$, mimicking EAS like events), and Lissajous images from an analogue oscilloscope (mimicking meteor-like events).

\begin{figure}
  \centering
  \includegraphics[width=12cm,height=8.4cm,clip]{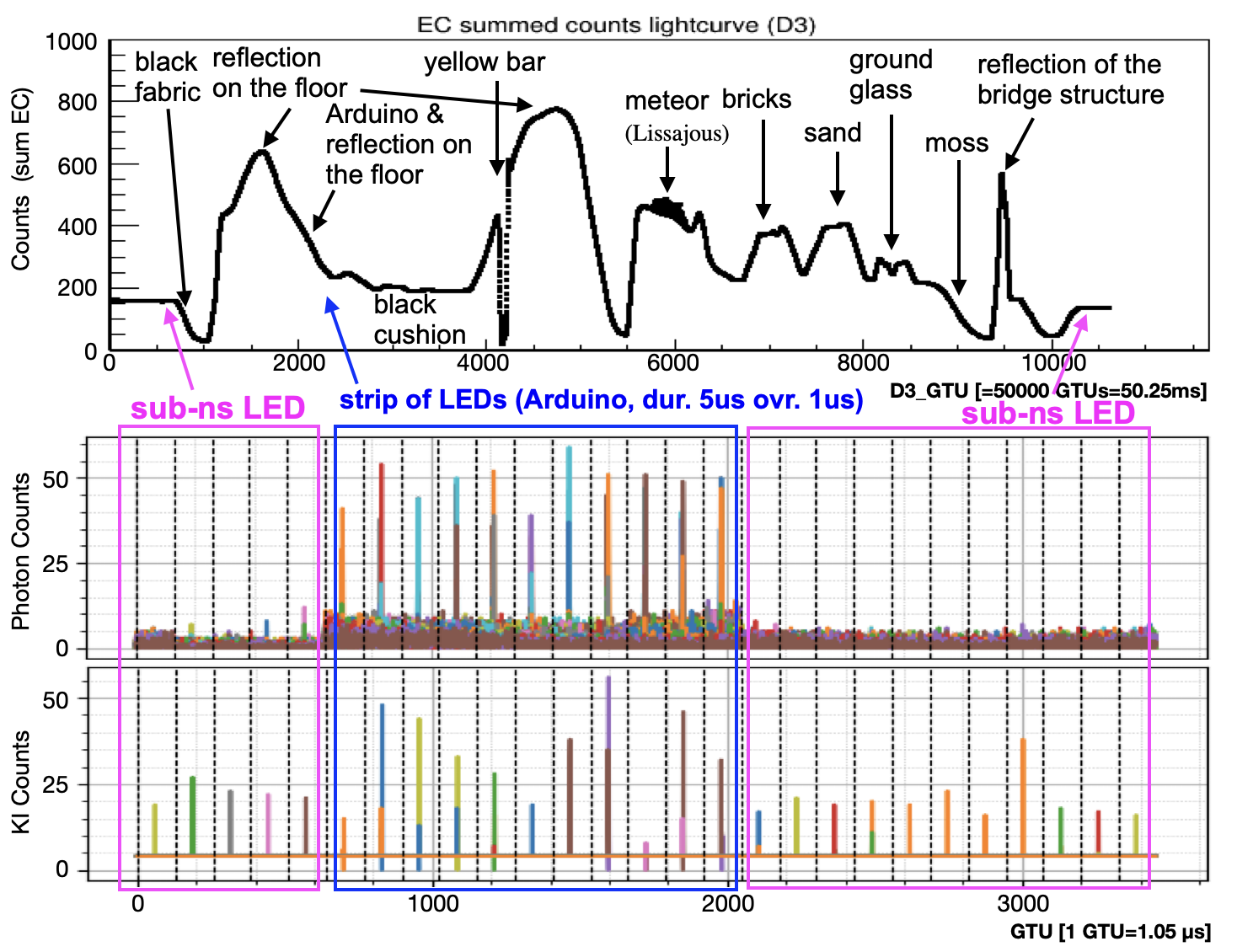}
  \caption{
    Top: Light curve showing the evolution of the photon counts averaged over 50ms periods, during a tank revolution ($\sim 9$ minutes in total). 
    Bottom: PC (middle) and KI counts (bottom) as a function of time (in GTUs), for events triggered by the KI trigger only (see text).
    \vspace{-0.3cm}    
  }
  \label{fig:TurLabD3}
  \end{figure}

The top panel of Fig.~\ref{fig:TurLabD3} shows the lightcurve measured by the EC during one entire rotation, with photon counts averaged over 50000 consecutive GTUs, i.e. 50~ms. The middle and bottom panels show respectively the photon counts per GTU and the corresponding KI values, as a function of GTU, for events triggered by the dedicated KI trigger during the periods indicated by the arrows. For this run, the Pixel\_THR and Ncounter parameters were set to 15 and 2 respectively.

The main result of these measurements is the confirmation of the KI trigger is working as expected, including in the presence of background light of different intensity, as well as during transitions from areas with different average photon counts. In particular, only short pulsed signals produced triggered events. It is also instructive to compare the triggered events obtained when the sub-ns LED pulses were in the field of view (periods in purple on the plot) and those when the Arduino-controlled LED strip were in the field of view (period in blue). In the case of the LED strip, the signals are long enough for large values of the photon counts to be measured by the ASIC, and the KI values are also large, because of the large integrated charge on the anodes. By contrast, in the case of the sub-ns light pulse, no significant signal is found in the photon counting mode, because of extreme pile-up, but the KI properly detects high values on very short time scales. This is indeed the intended behaviour, which confirms the relevance of this trigger mode for the operation of the instrument in flight.

\vspace{-0.3cm}
\section{Summary and Outlook}
\vspace{-0.3cm}

We have studied a new type of trigger for the JEM-EUSO FT instruments, intended to detect very short and intense flashes of light. This so-called ``KI trigger'' is based on the charge integration functionality of the SPACIROC-3 ASIC, whose characteristics were described both in the presence of intense continuous light or in response to short pulses of light. The tests performed with pulsed LED in Paris and in Turin, using the full detection chain of the EUSO-SPB2 ECs, demonstrated consistent behaviour of the KI channels in both modes, and confirmed the proper implementation of the KI trigger logic described in these proceedings. Finally, the tests performed at the TurLab facility allowed to demonstrate the performance and stability of the KI trigger in the presence of various levels of background light, including changing ones. Based on these measurements, the KI trigger was implemented and successfully operated in flight for the first time in the EUSO-SPB2 mission~\cite{ref:SPB2flight}.

\vspace{-0.3cm}
\section{Acknowledgments}
\vspace{-0.3cm}
The authors would like to acknowledge the support by NASA award 11-APRA-0058, 16-APROBES16-0023, 17-APRA17-0066, NNX17AJ82G, NNX13AH54G, 80NSSC18K0246,\\
80NSSC18K0473, 80NSSC19K0626, 80NSSC18K0464, 80NSSC22K1488, 80NSSC19K0627 and 80NSSC22K0426 and by National Science Centre in Poland grant n. 2017/27/B/ST9/02162.
This research used resources of the National Energy Research Scientific Computing Center (NERSC), a U.S. Department of Energy Office of Science User Facility operated under Contract No. DE-AC02-05CH11231.
This work was partially supported by the French space agency CNES.
We acknowledge the NASA Balloon Program Office and the Columbia Scientific Balloon Facility and staff for extensive support. 
We acknowledge the ASI-INFN agreement n. 2021-8-HH.0 and its amendments. 
We also acknowledge the invaluable contributions of the administrative and technical staffs at our home institutions.

%

%
%
%

\newpage
{\Large\bf Full Authors list: The JEM-EUSO Collaboration\\}

\begin{sloppypar}
{\small \noindent
S.~Abe$^{ff}$, 
J.H.~Adams Jr.$^{ld}$, 
D.~Allard$^{cb}$,
P.~Alldredge$^{ld}$,
R.~Aloisio$^{ep}$,
L.~Anchordoqui$^{le}$,
A.~Anzalone$^{ed,eh}$, 
E.~Arnone$^{ek,el}$,
M.~Bagheri$^{lh}$,
B.~Baret$^{cb}$
% ...
}
\end{sloppypar}
\vspace*{.3cm}

{ \footnotesize
\noindent
% ...
$^{cb}$ Universit\'e de Paris, CNRS, AstroParticule et Cosmologie, F-75013 Paris, France\\
% ...
$^{ed}$ Istituto Nazionale di Fisica Nucleare - Sezione di Catania, Italy\\
% ...
$^{eh}$ INAF - Istituto di Astrofisica Spaziale e Fisica Cosmica di Palermo, Italy\\
% ...
$^{ek}$ Istituto Nazionale di Fisica Nucleare - Sezione di Torino, Italy\\
$^{el}$ Dipartimento di Fisica, Universit\`a di Torino, Italy\\
% ...
$^{ep}$ Gran Sasso Science Institute, L'Aquila, Italy\\
% ...
$^{ff}$ Nihon University Chiyoda, Tokyo, Japan\\ 
% ...
$^{ld}$ University of Alabama in Huntsville, Huntsville, AL, USA\\
$^{le}$ Lehman College, City University of New York (CUNY), NY, USA\\
% ...
$^{lh}$ Georgia Institute of Technology, USA\\
% ...
}

\end{document}